\documentclass[%
superscriptaddress,
preprintnumbers,
tightenlines,
showpacs,
showkeys,
nofootinbib,
amsmath,
amssymb,
aps,
prd,
floatfix,
]{revtex4-1}

\usepackage{graphicx}
\usepackage{color}

\newcommand{\dof}{{\rm d.o.f.}}
\newcommand{\be}{\begin{equation}}
\newcommand{\ee}{\end{equation}}
\newcommand{\bea}{\begin{eqnarray}}
\newcommand{\eea}{\end{eqnarray}}
\newcommand{\non}{\nonumber}

\begin{document}
\title{Critical point phase transition for finite temperature 3-flavor QCD\\
  with non-perturbatively O($a$) improved Wilson fermions at $N_{\rm t}=10$}

\author{Xiao-Yong Jin}
\affiliation{Argonne Leadership Computing Facility, Argonne National Laboratory, Argonne, IL 60439, USA}

\author{Yoshinobu Kuramashi}
\affiliation{Center for Computational Sciences, University of Tsukuba, Tsukuba, Ibaraki 305-8577, Japan}
\affiliation{RIKEN Advanced Institute for Computational Science, Kobe, Hyogo 650-0047, Japan}

\author{Yoshifumi~Nakamura}
\affiliation{RIKEN Advanced Institute for Computational Science, Kobe, Hyogo 650-0047, Japan}
\affiliation{Graduate School of System Informatics, Department of Computational Sciences, Kobe University, Kobe, Hyogo 657-8501, Japan}

\author{Shinji Takeda}\email[]{takeda@hep.s.kanazawa-u.ac.jp}
\affiliation{Institute of Physics, Kanazawa University, Kanazawa 920-1192, Japan}
\affiliation{RIKEN Advanced Institute for Computational Science, Kobe, Hyogo 650-0047, Japan}

\author{Akira Ukawa}
\affiliation{RIKEN Advanced Institute for Computational Science, Kobe, Hyogo 650-0047, Japan}

\date{\today}
\begin{abstract}
We study the finite temperature phase structure for three-flavor QCD with a
focus on locating the critical point which separates crossover and first order phase transition region
in the chiral regime of the Columbia plot.
In this study, we employ the Iwasaki gauge action and the non-perturvatively O($a$) improved Wilson-Clover fermion action.
We discuss the finite size scaling analysis including the mixing of magnetization-like and energy-like observables. 
We carry out the continuum extrapolation of the critical point using newly generated data
at $N_{\rm t}=8$, $10$ and estimate the upper bound of the critical pseudo-scalar meson mass
$m_{\rm PS,E}
\lesssim
170
{\rm MeV}
$
and the critical temperature
$
T_{\rm E}
=
134(3)
{\rm MeV}
$.
Our estimate of the upper bound is derived from the existence of the critical point as an edge of the 1st order phase transition while
that of the staggered-type fermions with smearing is based on its absence.
\end{abstract}


\preprint{UTHEP-703, UTCCS-P-103, KANAZAWA-17-05}

\maketitle

\section{Introduction}
The nature of finite temperature transition in QCD varies depending on the quark masses.
A pictorial representation is often given as the Columbia plot \cite{Brown:1990ev,Pisarski:1983ms,Gavin:1993yk} whose axes
are usually taken to be  the up-down and strange quark masses.
See reviews \cite{Schmidt:2017bjt,Ding:2017giu} for a current status of the QCD phase structure with the finite temperature and quark number density.
In this paper we restrict ourselves to the case of zero quark number density.

There are two longstanding issues on the Columbia plot, namely
the location of the critical line which separates the first  order phase transition region from the cross over region,
and the universality class of the critical line. 
Studies with  the standard staggered fermion action \cite{Aoki:1998gia,Karsch:2001nf,deForcrand:2006pv,Smith:2011pm}
successfully located the critical point along the flavor symmetric line ($N_{\rm f}=3$).  
It was subsequently found that the first order region rapidly shrinks towards the continuum limit \cite{deForcrand:2007rq}.
Further studies with  staggered fermions with smearing techniques \cite{Endrodi:2007gc,Ding:2011du,Bazavov:2017xul}
could not even detect a critical point, perhaps due to the possibility that the critical quark mass is so small that current computational resources cannot access it.

On the other hand, the pioneering Wilson-type fermion study in Ref.~\cite{Iwasaki:1996zt} reported a relatively heavy critical mass.
Our recent study \cite{Jin:2014hea}, while confirming a large value for coarse lattice spacings, suggested that the critical mass appears to be smaller for finer lattice spacings.
This implies  that removal of  scaling violation is crucial for Wilson-type fermion action as well.
In Ref.~\cite{Jin:2014hea}, we computed the critical point for $N_{\rm f}=3$ QCD at temporal lattice sizes $N_{\rm t}=4$, $6$ and $8$.
In order to take the continuum limit more reliably, we have recently started large scale simulations at $N_{\rm t}=10$
and preliminary results were already reported in the previous lattice conferences \cite{nakamuralattice2015,Takeda:2016vfj}.
In this paper, we finalize the analysis including the new data, which consists of one additional $\beta$ value with $N_{\rm t}=8$ and totally new data set of $N_{\rm t}=10$,
and examine the continuum limit with the added data. 

Concerning the issue with the universality class along the
critical line,
we observed in Ref.~\cite{Jin:2014hea,nakamuralattice2015} that for $N_{\rm
  t}=8$ and $10$ the values of the kurtosis from different
volumes intersect at a point away from three-dimensional Z$_2$ universality class,
in contrast to the situation with $N_{\rm t}=4$ and $6$ where they are consistent.
We address this issue by noting that bare lattice observables generally are  mixtures of magnetization-like and energy-like operators; this should be taken into account in finite size scaling analyses.

The rest of the paper is organized as follows.
In section \ref{sec:scaling}, we review the kurtosis intersection analysis
and then discuss the finite size correction for kurtosis of an observable,
which has a non-trivial overlap with the energy-like operator, around the critical point.
After describing the simulation setup in section \ref{sec:setup}, we locate the critical point by applying the new fitting formulae
in kurtosis intersection analysis and then take the continuum limit of the critical point in section \ref{sec:results}.
Our conclusions are summarized in section \ref{sec:summary}.
Results of zero temperature simulations for scale setting are summarized in appendix \ref{sec:scale}.

\section{Scaling analysis for general observable}
\label{sec:scaling}
In this section we review the standard kurtosis intersection formula and
derive a new formula which incorporates the finite volume effect for kurtosis of a general observable
which is a mixture of energy-like and magnetization-like operator 
around the critical point.
The mixed observable analysis was originally discussed and demonstrated in Ref.~\cite{Karsch:2001nf}, where
first of all the magnetization part is extracted by using some observables
and then the kurtosis intersection analysis is applied to the magnetization-dominated observable.
Here, however, we consider a general observable without purifying the magnetization part
and derive a formula of the kurtosis with correction term originated from the energy-like operator part.
The derived formula will be used in the subsequent analysis.

In a scaling analysis,
the relevant parameters are reduced temperature $t$, external magnetic field $h$ and the inverse of the linear lattice size $L^{-1}$.
According to finite size scaling theory, under scaling by a factor $b$, 
the free energy (not free energy density) scales as follows up to analytic terms:
\be
F(t,h,L^{-1})=F(tb^{y_{\rm t}},hb^{y_{\rm h}},L^{-1}b),
\label{eqn:Frescale}
\ee
where $y_{\rm t}$ and $y_{\rm h}$ are the exponent for the temperature and the magnetic field, respectively.
Setting $b=L$, 
the scaling relation of the free energy is given by
\be
F(t,h,L^{-1})=F(tL^{y_{\rm t}},hL^{y_{\rm h}},1).
\ee
In the following,
we use the notation and abbreviation below
\be
F(tL^{y_{\rm t}},0,1)
=
F(tL^{y_{\rm t}}),
\hspace{10mm}
\frac{\partial^n}{\partial t^{n}}
\frac{\partial^m}{\partial h^{m}}
F(t,h,L^{-1})
=
F^{(nm)}(t,h,L^{-1}).
\ee

As an explicit and well known example, first we consider the purely magnetic observable ${\cal M}$,
whose moments can be obtained by applying the derivative in terms of $h$ to the free energy
\be
{\cal M}\rightarrow \frac{\partial}{\partial h}.
\ee
The susceptibility for ${\cal M}$ at $h=0$ is given by
\be
\chi_{\cal M}(t,0,L^{-1})
=
L^{-d}
\left.
\frac{\partial^2 F(t,h,L^{-1})}{\partial h^2}
\right|_{h=0}
=
L^{-d+2y_h}F^{(02)}(tL^{y_t},0,1)
=
L^{-d+2y_h}F^{(02)}(tL^{y_t}).
\ee
where $d$ is the dimension of the system.
As well known, the susceptibility at $t=0$ scales as (using $-d+2y_h=\gamma/\nu$)
\be
\chi_{\cal M}(0,0,L^{-1})
\propto
L^{\gamma/\nu}.
\label{eqn:chi}
\ee
The kurtosis for ${\cal M}$ at $h=0$ is given by
\be
K_{\cal M}(t,0,L^{-1})
=
\frac
{\left.F^{(04)}(t,h,L^{-1})\right|_{h=0}}
{\left[\left.F^{(02)}(t,h,L^{-1})\right|_{h=0}\right]^2}
=
\frac
{L^{4y_{\rm h}}F^{(04)}(tL^{y_{\rm t}})}
{\left[L^{2y_{\rm h}}F^{(02)}(tL^{y_{\rm t}})\right]^2}
=
\frac
{F^{(04)}(tL^{y_{\rm t}})}
{\left[F^{(02)}(tL^{y_{\rm t}})\right]^2}.
\ee
At a critical point, the kurtosis is independent of the volume.
For small $tL^{y_{\rm t}}$,
one can expand
\be
\label{eqn:km}
K_{\cal M}(t,0,L^{-1})
=
\frac
{F^{(04)}(0)}
{\left[F^{(02)}(0)\right]^2}
+
c_K
tL^{1/\nu}
+...,
\ee
where we have used $y_{\rm t}=1/\nu$.
This is the well known formula for the kurtosis intersection analysis.

For a general observable ${\cal O}$ which is a mixture of energy ${\cal E}$ and magnetization ${\cal M}$,
\be
{\cal O}=
c_{\rm M}
{\cal M}
+
c_{\rm E}
{\cal E}
\rightarrow
c_{\rm M}
\frac{\partial}{\partial h}
+
c_{\rm E}
\frac{\partial}{\partial t},
\ee
the susceptibility and kurtosis of ${\cal O}$ at $h=0$ are given by
\bea
\chi_{\cal O}(t,0,L^{-1})
&=&
L^{-d}
\left.
\left(
c_M
\frac{\partial}{\partial h}
+
c_E
\frac{\partial}{\partial t}
\right)^2
F(t,h,L^{-1})
\right|_{h=0}
\non\\
&=&
L^{-d+2y_h} c_M^2
\left[
F^{(02)}(tL^{y_t})
+
2\frac{c_E}{c_M} L^{y_t-y_h} F^{(11)}(tL^{y_t})
+
O(L^{2(y_t-y_h)})
\right],
\label{eqn:chiO}
\\
K_{\cal O}(t,0,L^{-1})
&=&
\frac{
\left.
\left(
c_{\rm M}
\frac{\partial}{\partial h}
+
c_{\rm E}
\frac{\partial}{\partial t}
\right)^4
F(t,h,L^{-1})
\right|_{h=0}
}{
\left[
\left.
\left(
c_{\rm M}
\frac{\partial}{\partial h}
+
c_{\rm E}
\frac{\partial}{\partial t}
\right)^2
F(t,h,L^{-1})
\right|_{h=0}
\right]^2
}
\non\\
&=&
\frac{F^{(04)}(tL^{y_{\rm t}})}{F^{(02)}(tL^{y_{\rm t}})^2}
\left[
1+
\frac{4c_{\rm E}}{c_{\rm M}}
L^{y_{\rm t}-y_{\rm h}}
\left(
\frac{F^{(13)}(tL^{y_{\rm t}})}{F^{(04)}(tL^{y_{\rm t}})}
-
\frac{F^{(11)}(tL^{y_{\rm t}})}{F^{(02)}(tL^{y_{\rm t}})}
\right)
+
O(L^{2(y_{\rm t}-y_{\rm h})})
\right].
\label{eqn:kurtosisO}
\eea
Thus, even when setting $t=0$,
the correction term of $O(c_{\rm E}L^{y_{\rm t}-y_{\rm h}}/c_{\rm M})$ remains\footnote{
For the kurtosis, there is another correction term originated from the irrelevant scaling field $N_{\rm s}^{-1/\nu-\omega}$.
The value of $\omega$ for three-dimensional Z$_2$ universality class is $0.83...$ and the magnitude of the correction term is similar to that of the mixing.
In numerical analysis it is hard to disentangle them.
Therefore in this paper, we deal with only the dominant mixing correction term and just ignore the irrelevant contribution.
We thank de Forcrand for reminding us this issue.
}.
In particular, the correction term alters the value of kurtosis at the critical point.
The difference of the exponents $y_{\rm t}-y_{\rm h}$ is usually negative for various universality classes, {\it viz.}
\be
y_{\rm t}-y_{\rm h}
=
\frac{1}{2\nu}(\alpha-\gamma)
=
\left\{
\begin{array}{lll}
\frac{1}{2\cdot1}(0-7/4)&=-7/8 &\mbox{ : 2D Ising},\\
\frac{1}{2\cdot0.630}(0.110-1.237)&=-0.894 &\mbox{ : 3D Ising},\\
\frac{1}{2\cdot0.67}(-0.01-1.32)&=-0.993 &\mbox{ : 3D O(2)},\\
\frac{1}{2\cdot0.75}(-0.25-1.47)&=-1.15 &\mbox{ : 3D O(4)}.
\end{array}
\right.
\ee
Therefore, such a correction would be irrelevant in the large volume limit.  
However, at finite volumes,  the value of kurtosis at $t=0$ has a volume dependence;  
the kurtosis for various volumes would not cross at a single point.

\section{Setup and methods}
\label{sec:setup}

We employ the Iwasaki gauge action \cite{Iwasaki:2011np} and non-perturvatively O($a$) improved Wilson-Clover fermion action \cite{Aoki:2005et}
to carry out the finite temperature $N_{\rm f}=3$ QCD simulation.
The temporal lattice size we newly report here is a part of $N_{\rm t}=8$ and all $N_{\rm t}=10$ data.
For $N_{\rm t}=8$ and $10$, the spatial lattice size is varied over $N_{\rm s}=16$, $20$, $24$ and $28$ to carry out  finite size scaling.
BQCD code \cite{Nakamura:2010qh} implementing the RHMC algorithm \cite{Clark:2006fx} is used to generate gauge configurations,
with the acceptance rate tuned to be around 80\%.
We store configurations at every 10th trajectory for observable measurements. 
Since the three dynamical quarks are all degenerate, we have only one hopping parameter $\kappa$.
Some values  of the parameter $\beta$ are selected,  and  $\kappa$ is adjusted
to search for a transition point at each $\beta$.
See Table \ref{tab:ensemble} for the parameter sets and their statistics.
The new $\beta$ value for $N_{\rm t}=8$, which was absent in \cite{Jin:2014hea}, is $\beta=1.745$.

The details of our analysis method can be found in our previous studies
\cite{Jin:2014hea,Kuramashi:2016kpb}, and we summarize it in the
following.

We use the naive chiral condensate as a probe to study the phase structure.
We measure higher moments of the chiral condensate up to the fourth order
to calculate the susceptibility, the skewness, and the kurtosis
equivalent to the Binder cumulant up to an additional constant.
In order to determine the transition point, 
we use the peak position of the susceptibility and verify that it
coincides with the zero of the skewness.
The kurtosis is used to locate the critical point through the
intersection analysis \cite{Karsch:2001nf} with our extension
discussed in the previous section.

We combine several ensembles, which share common parameter values except for $\kappa$,
by the multi-ensemble reweighting \cite{Ferrenberg:1988yz} in $\kappa$ for interpolating the moments. We do not apply $\beta$-reweighting.  
To calculate the reweighting factor given by the ratio of fermion determinants at different $\kappa$ values,
we use an expansion of the logarithm of the determinant \cite{Kuramashi:2016kpb}.
For the computation of the observable part in the reweighting procedure,
we need to evaluate quark propagators at continuously many points of $\kappa$.
We adopt an expansion form for the moments which allows us to evaluate the moments at continuously many points at a relatively low cost.
The multi-ensemble reweighting is applied to the data at $N_{\rm t}=8$ and $10$ as well as the old $N_{\rm t}=4, 6$ data without
adding new data set.
For all values of $N_{\rm t}$, our parameter sets satisfy $m_{\rm PS} L\gtrsim4$ where $m_{\rm PS}$ is the pseudo-scalar meson mass.

\begin{table}[!hbt]
\caption{Simulation parameters and the number of configurations for $N_{\rm t}=8$ and $10$.}
\label{tab:ensemble}%
\begin{ruledtabular}
\begin{tabular}{lll|r|r|r|r}
$N_{\rm t}$& $ \beta $ & $\kappa$ & $N_{\rm s}=16$ & $N_{\rm s}=20$ & $N_{\rm s}=24$ & $N_{\rm s}=28$ \\
\hline
8&$ 1.745 $&$ 0.140371 $&$ 5600 $&$ 3050 $&$  850 $&$  400$ \\
 &$       $&$ 0.140380 $&$ 6170 $&$ 7890 $&$    - $&$    -$ \\
 &$       $&$ 0.140382 $&$    - $&$    - $&$11200 $&$11830$ \\
 &$       $&$ 0.140384 $&$    - $&$    - $&$    - $&$ 6670$ \\
 &$       $&$ 0.140385 $&$    - $&$ 9910 $&$15700 $&$    -$ \\
 &$       $&$ 0.140393 $&$14570 $&$    - $&$    - $&$    -$ \\
\cline{2-7}
 &$1.74995$&$ 0.140240 $&$15498 $&$10700 $&$ 9700 $&$ 6580$ \\
\cline{2-7}
 &$1.76019$&$ 0.139950 $&$16650 $&$11230 $&$10560 $&$    -$ \\
\hline
10&$ 1.77 $&$ 0.139800 $&$  640 $&$    - $&$    - $&$    -$ \\
  &$      $&$ 0.139820 $&$ 1620 $&$    - $&$    - $&$    -$ \\
  &$      $&$ 0.139830 $&$ 1520 $&$    - $&$    - $&$    -$ \\
  &$      $&$ 0.139850 $&$ 3510 $&$ 2000 $&$    - $&$    -$ \\
  &$      $&$ 0.139855 $&$    - $&$ 2410 $&$  830 $&$    -$ \\
  &$      $&$ 0.139857 $&$    - $&$    - $&$  380 $&$    -$ \\
  &$      $&$ 0.139858 $&$    - $&$    - $&$  710 $&$    -$ \\
  &$      $&$ 0.139860 $&$    - $&$ 2500 $&$  630 $&$    -$ \\
  &$      $&$ 0.139870 $&$ 3590 $&$    - $&$    - $&$    -$ \\
  &$      $&$ 0.139900 $&$  760 $&$    - $&$    - $&$    -$ \\
\cline{2-7}
  &$ 1.78 $&$ 0.139550 $&$ 1220 $&$    - $&$    - $&$    -$ \\
  &$      $&$ 0.139560 $&$ 1520 $&$    - $&$    - $&$    -$ \\
  &$      $&$ 0.139580 $&$ 2720 $&$    - $&$    - $&$    -$ \\
  &$      $&$ 0.139600 $&$ 2870 $&$    - $&$    - $&$    -$ \\
  &$      $&$ 0.139610 $&$ 2640 $&$ 1640 $&$  720 $&$    -$ \\
  &$      $&$ 0.139615 $&$    - $&$ 3810 $&$ 2320 $&$ 1610$ \\
  &$      $&$ 0.139620 $&$ 2460 $&$ 4400 $&$ 2360 $&$ 1110$ \\
  &$      $&$ 0.139625 $&$    - $&$  690 $&$  550 $&$    -$ \\
  &$      $&$ 0.139630 $&$    - $&$  490 $&$    - $&$    -$ \\
  &$      $&$ 0.139650 $&$ 1790 $&$    - $&$    - $&$    -$ \\
\cline{2-7}
  &$ 1.79 $&$ 0.139300 $&$ 2760 $&$    - $&$    - $&$    -$ \\
  &$      $&$ 0.139325 $&$ 2710 $&$    - $&$    - $&$    -$ \\
  &$      $&$ 0.139340 $&$    - $&$ 2660 $&$  730 $&$    -$ \\
  &$      $&$ 0.139350 $&$ 3270 $&$ 2380 $&$  960 $&$    -$ \\
  &$      $&$ 0.139400 $&$ 3460 $&$    - $&$    - $&$    -$ \\
\end{tabular}
\end{ruledtabular}
\end{table}

\section{Results}
\label{sec:results}

\subsection{Moments and location of the transition point}

\begin{figure}[t!]
\begin{center}
\begin{tabular}{cc}
\scalebox{0.9}{\includegraphics{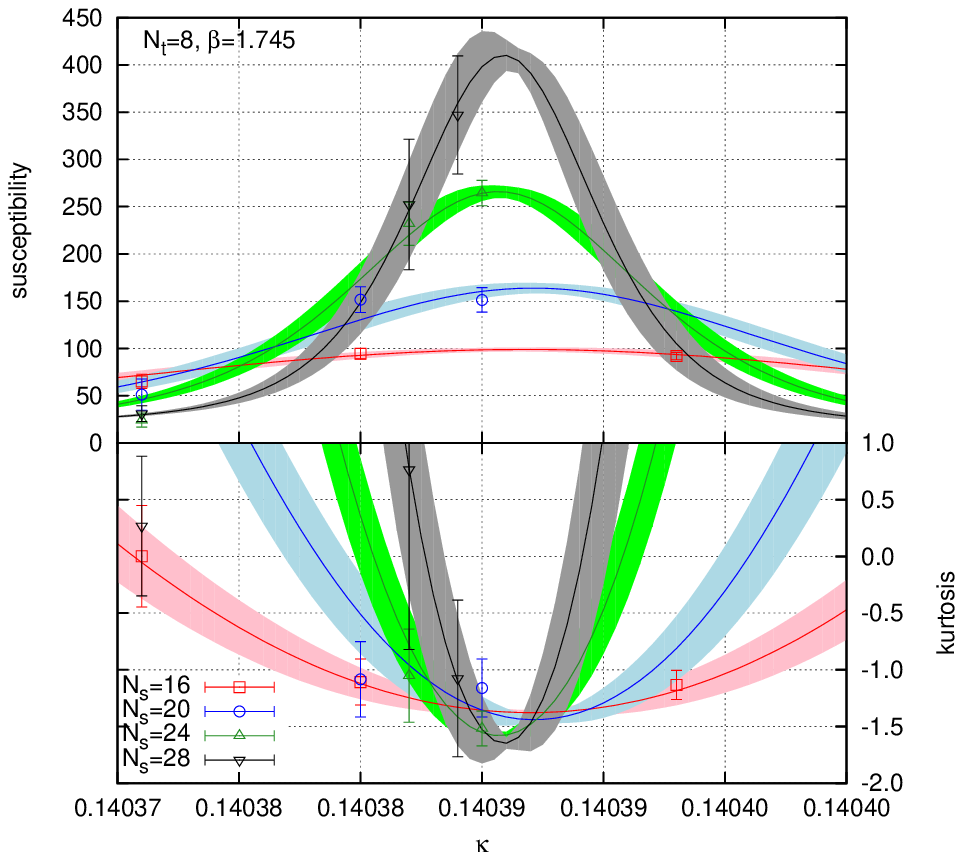}}
&
\hspace{-30mm}
\scalebox{0.9}{\includegraphics{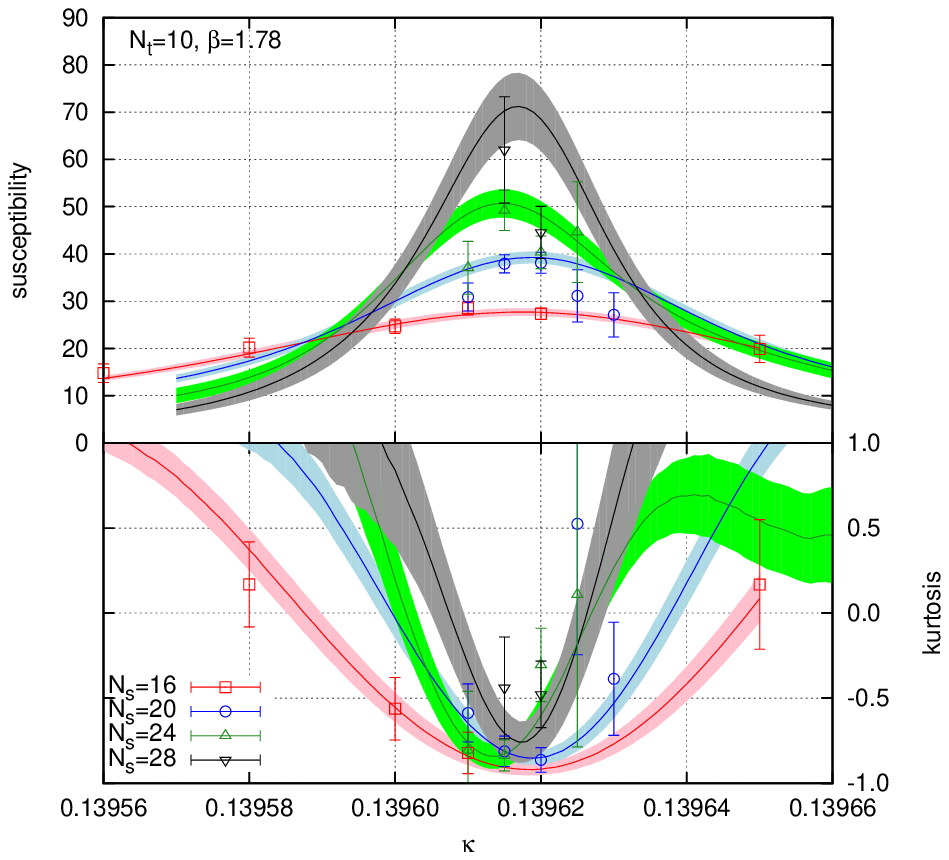}}
\end{tabular}
\end{center}
\caption{The susceptibility (upper half) and kurtosis (lower half) of chiral condensate as a function $\kappa$
with several spatial sizes, $N_{\rm s}=16-28$.
The left panel is for $(N_{\rm t},\beta)=(8,1.745)$ and the right is for $(N_{\rm t},\beta)=(10,1.78)$.
The raw data points (as symbols) as well as the multi-ensemble reweighting (1-$\sigma$ band) are plotted.
}
\label{fig:moment}
\end{figure}

\begin{figure}[h!]
\begin{center}
\begin{tabular}{c}
\scalebox{1.}{\includegraphics{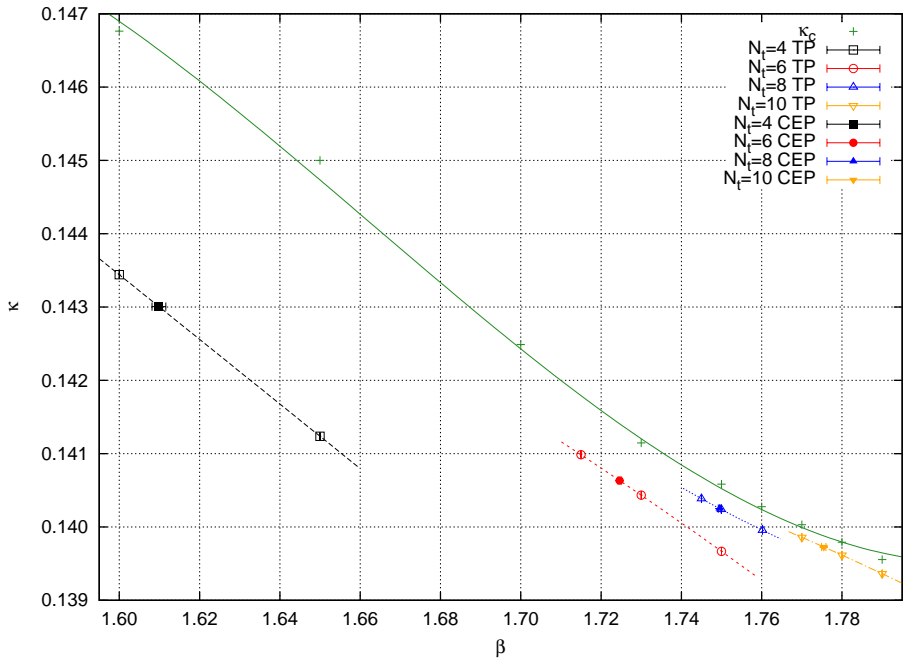}}
\end{tabular}
\end{center}
\caption{
Phase diagram for bare parameter space $(\beta,\kappa)$ at $N_{\rm t}=4$, $6$, $8$ and $10$.
The open symbols represent a transition/crossover point while the
filled symbols are the critical end points
determined by the kurtosis intersection with new formula.
On the transition line, the left (right) hand side of an
critical end point
is the first order phase transition (crossover) side.
Polynomial interpolation is used to determine the phase transition line.
$\kappa_{\rm c}$ is the pseudo-scalar massless point with $N_{\rm f}=3$ at the zero temperature.
}
\label{fig:phase_diagram}
\end{figure}

As an illustration of new data, we show the susceptibility and the kurtosis of the chiral condensate
for $(N_{\rm t},\beta)=(8,1.745)$ and $(N_{\rm t},\beta)=(10,1.78)$ in Fig.~\ref{fig:moment}
together with the $\kappa$-reweighting results.
From the peak position of the susceptibility, we extract the transition points.
The thermodynamic limit of the transition point is taken by using a fitting form with an inverse spatial volume correction term.
The resulting phase diagram in the bare parameter space is summarized in Fig.~\ref{fig:phase_diagram}.
Polynomial interpolation is used to determine the phase transition line.

\subsection{Kurtosis analysis}

\begin{figure}[h!]
\begin{center}
\begin{tabular}{cc}
\scalebox{1.}{\includegraphics{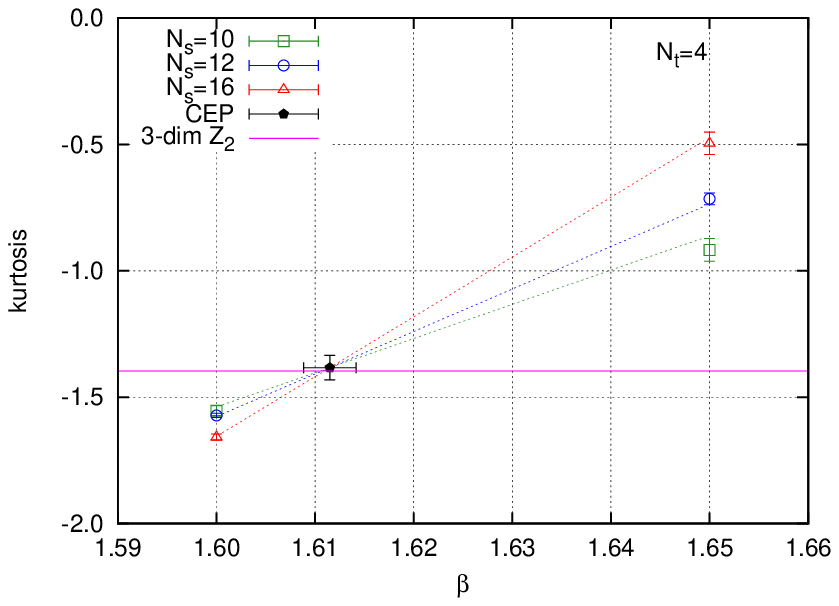}}
&
\scalebox{1.}{\includegraphics{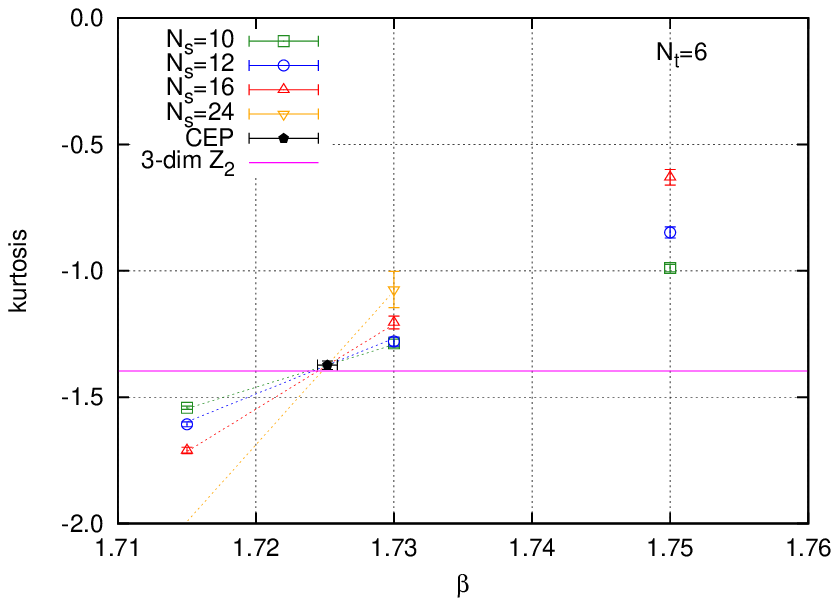}}
\\
\scalebox{1.}{\includegraphics{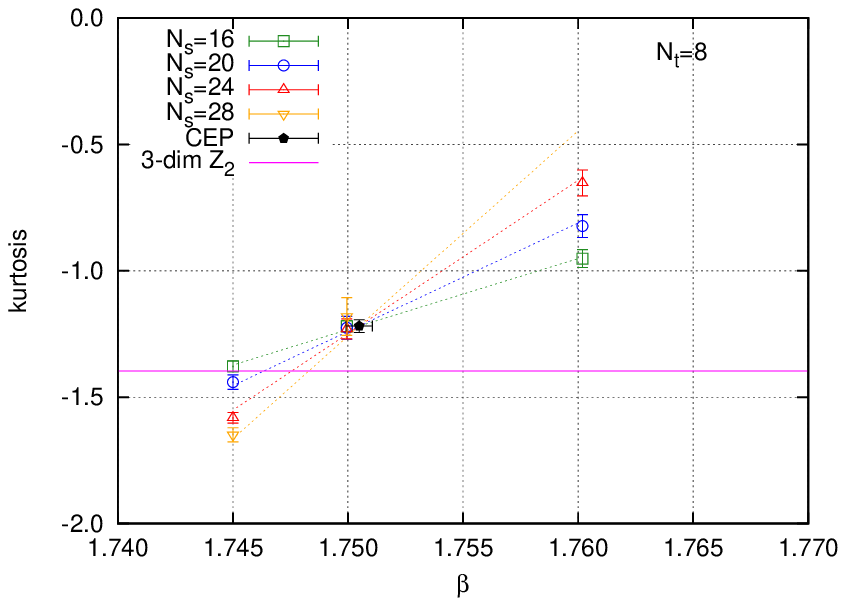}}
&
\scalebox{1.}{\includegraphics{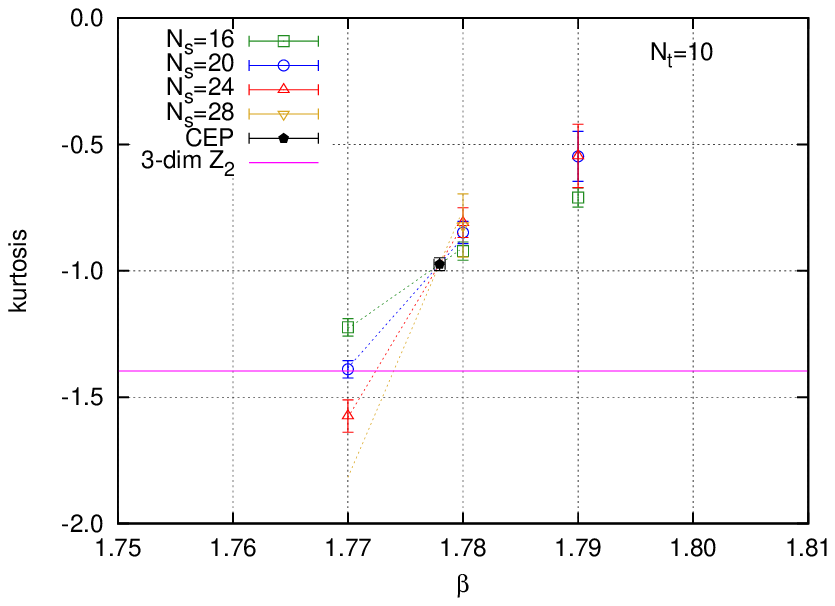}}
\end{tabular}
\end{center}
\caption{
Kurtosis intersection for chiral condensate at $N_{\rm t}=4$, $6$, $8$ and $10$.
The simple fitting form: $K=K_{\rm E}+AN_{\rm s}^{1/\nu}(\beta-\beta_{\rm E})$ is also drawn in the figure.
For $N_{\rm t}=4$ and $6$, the values of kurtosis at the crossing point (black pentagon) is consistent with the three-dimensional Z$_2$ universality class, while
for $N_{\rm t}=8$ and $10$, obviously it is not consistent.
The conclusion of this paper is based on the fitting form
  with a correction term.  See text for details.
}
\label{fig:krt}
\end{figure}

The minimum of kurtosis at each $(N_{\rm s},\beta)$ is plotted in Fig.~\ref{fig:krt}
to perform  kurtosis intersection analysis at $N_{\rm t}=4$, $6$, $8$ and $10$.
Although $N_{\rm t}=4$ and $6$ results clearly show that the critical universality class is consistent with the 3D Z$_2$ universality class,
for $N_{\rm t}=8$ and $10$ an analysis using the conventional formula
as eq.~(\ref{eqn:km})
leads to a value for $K_{\rm E}$ which is substantially larger than that for the universality class.
In this situation we attempt a modified fitting form  which
incorporates the correction term in eq.(\ref{eqn:kurtosisO})
associated with the contribution of the energy-like observable given by 
\be
K
=
\left[
K_{\rm E}
+
AN_{\rm s}^{1/\nu}(\beta-\beta_{\rm E})
\right]
(1+BN_{\rm s}^{y_{\rm t}-y_{\rm h}}),
\label{eqn:new}
\ee
where we have two additional parameters $B$ and $y_{\rm t}-y_{\rm h}$.
We have tried three fits. Fit-1 has no correction term ($B=y_{\rm t}-y_{\rm h}=0$) and all other parameters are used as fit parameter.
Fit-2 also neglects the correction term assuming the 3D Z$_2$ universality class for $K_{\rm E}$ and $\nu$.
Fit-3 includes the correction term assuming the 3D Z$_2$ universality class for $K_{\rm E}$, $\nu$ and $y_{\rm t}-y_{\rm h}$.
The fit results are summarized in Table \ref{tab:CEP}.

For $N_{\rm t}=4$ and $6$, the parameter $B$ in the Fit-3 is consistent with zero
and all other fitting parameters of all fitting forms are consistent with each other.
Thus we conclude that the new correction term is negligible and universality class is  consistent with 3D Z$_2$ for $N_{\rm t}=4$ and $6$.

For $N_{\rm t}=8$ and $10$,
the assumption of Z$_2$ universality class is unlikely to hold without the new correction term 
since the Fit-2, which assumes the Z$_2$ values for $K_{\rm E}$ and $\nu$,  has  large $\chi^2/{\rm d.o.f.}$
On the other hand, with the Fit-3 assuming Z$_2$ but including
the correction term from mixing of magnetization and energy
terms, we observe a reasonable $\chi^2/{\rm d.o.f.}<1$.  
The magnitude of the correction term,
$BN_{\rm s}^{y_{\rm t}-y_{\rm h}}$,
is reasonably small, of order 10\%.  
This suggests that $N_{\rm t}=8$ and $10$ results are  consistent with the 3D Z$_2$ universality class
if one includes the  correction term.

In the following we adopt the critical point $\beta_{\rm E}$ determined by the Fit-3 (assuming 3D Z$_2$ universality class with the correction term).
The corresponding critical value of $\kappa$, that is, $\kappa_{\rm E}$ is estimated by an interpolated transition line as in Fig.~\ref{fig:phase_diagram}
where the critical point in the bare parameter space $(\beta,\kappa)$ is shown.

\begin{table}[t!]
\caption{
Fit results for kurtosis intersection with fitting form in eq.(\ref{eqn:new}).
See text for the definition of Fit-1, 2 and 3.
A value without error bar means that the corresponding fit parameter is fixed to the given value during the fit.
For the 3D Z$_2$ universality class, the expected values of the parameter are
$K_{\rm E}=-1.396$, $\nu=0.630$ and $y_{\rm t}-y_{\rm h}=-0.894$ respectively.
Using the value of $\beta_{\rm E}$ as an input,
$\kappa_{\rm E}$ is obtained from an interpolation formula of the transition line in Fig.~\ref{fig:phase_diagram}.
}
\label{tab:CEP}
\begin{ruledtabular}
\begin{tabular}{rrllllllrr}
$N_{\rm t}$
&
Fit
&
$\beta_{\rm E}$
&
$\kappa_{\rm E}$
&
$K_{\rm E}$
&
$\nu$
&
$A$
&
$B$
&
$y_{\rm t}-y_{\rm h}$
&
$\chi^2/{\rm d.o.f.}$
\\
\hline
4&
$1$&$1.6115 ( 26 )$&$0.1429337 ( 13 )$&$-1.383 ( 48 )$&$0.84 ( 13 )$&$0.88 ( 42 )$&$\times$&$\times$&$1.75$\\
&
$2$&$1.61065 ( 61 )$&$0.1429713 ( 13 )$&$-1.396$&$0.63$&$0.313 ( 12 )$&$\times$&$\times$&$3.05$\\
&
$3$&$1.6099 ( 17 )$&$0.1430048 ( 13 )$&$-1.396$&$0.63$&$0.311 ( 14 )$&$0.10 ( 21 )$&$-0.894$&$3.77$\\
\hline
6&
$1$&$1.72518 ( 71 )$&$0.1406129 ( 14 )$&$-1.373 ( 17 )$&$0.683 ( 54 )$&$0.58 ( 17 )$&$\times$&$\times$&$0.68$\\
&
$2$&$1.72431 ( 24 )$&$0.1406451 ( 14 )$&$-1.396$&$0.63$&$0.418 ( 11 )$&$\times$&$\times$&$0.70$\\
&
$3$&$1.72462 ( 40 )$&$0.1406334 ( 14 )$&$-1.396$&$0.63$&$0.422 ( 12 )$&$-0.052 ( 52 )$&$-0.894$&$0.70$\\
\hline
8&
$1$&$1.75049 ( 57 )$&$0.1402234 ( 11 )$&$-1.219 ( 25 )$&$0.527 ( 55 )$&$0.146 ( 88 )$&$\times$&$\times$&$0.73$\\
&
$2$&$1.74721 ( 42 )$&$0.14031921 ( 76 )$&$-1.396$&$0.63$&$0.404 ( 36 )$&$\times$&$\times$&$5.99$\\
&
$3$&$1.74953 ( 33 )$&$0.1402512 ( 10 )$&$-1.396$&$0.63$&$0.414 ( 13 )$&$-1.33 ( 15 )$&$-0.894$&$0.73$\\
\hline
10&
$1$&$1.77796 ( 48 )$&$0.1396661 ( 17 )$&$-0.974 ( 25 )$&$0.466 ( 45 )$&$0.084 ( 52 )$&$\times$&$\times$&$0.22$\\
&
$2$&$1.7694 ( 16 )$&$0.1398724 ( 22 )$&$-1.396$&$0.63$&$0.421 ( 95 )$&$\times$&$\times$&$10.03$\\
&
$3$&$1.77545 ( 53 )$&$0.1397274 ( 17 )$&$-1.396$&$0.63$&$0.559 ( 29 )$&$-2.97 ( 25 )$&$-0.894$&$0.43$\\
\end{tabular}
\end{ruledtabular}
\end{table}

\subsection{Cross check using exponent of the susceptibility peak height}
\begin{figure}[t]
\begin{center}
\begin{tabular}{c}
\scalebox{1}{\includegraphics{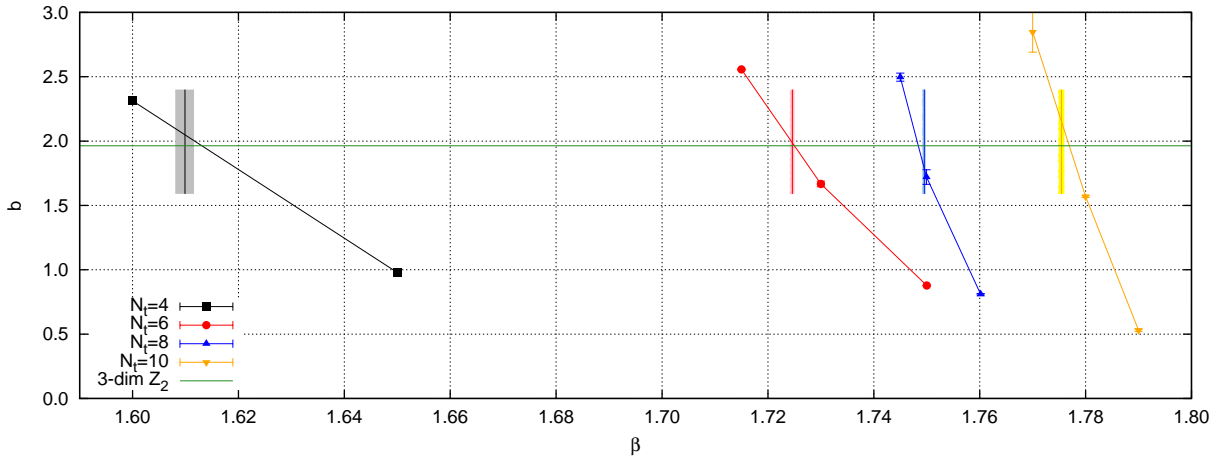}}
\end{tabular}
\end{center}
\caption{
Exponents of the susceptibility peak height along the transition line projected on $\beta$ value for $N_{\rm t}=4$, $6$ ,$8$ and $10$.
The line connecting the data points is to guide readers' eyes.
The point where the line for each $N_{\rm t}$ intersects the (green) horizontal line is an estimate of the critical point assuming the Z$_2$ universality class.
On the other hand, the shaded areas represent the critical $\beta$ determined by the kurtosis intersection analysis.
}
\label{fig:exponent}
\end{figure}

For a cross check of the location and the universality class of the critical point,
we investigate the scaling of the susceptibility peak height for the chiral condensate,
\begin{equation}
\chi_{\max}
\propto
(N_{\rm s})^b.
\label{eqn:chimax}
\end{equation}
At a critical point, the exponent should be $b=\gamma/\nu$ as in eq.(\ref{eqn:chi}).
For a general observable, the dominant part shows the same
scaling as in eq.(\ref{eqn:chimax}) while a correction term as in
eq.(\ref{eqn:chiO}) remains even at the critical point.
For a qualitative verification we neglect the correction
  term and extract the exponent $b$ with a log-linear fit.
The resulting exponent $b$ is plotted in Fig.~\ref{fig:exponent} along the transition line projected on $\beta$.
Assuming the Z$_2$ universality class provides an estimation of the critical point of $\beta$,
we confirm that it
is consistent with that of the kurtosis intersection.
This cross check assures that our analysis is working well.

\subsection{Continuum extrapolation of critical pseudoscalar meson  mass and critical temperature}

\begin{figure}[t!]
\begin{center}
\begin{tabular}{cc}
\scalebox{0.7}{\includegraphics{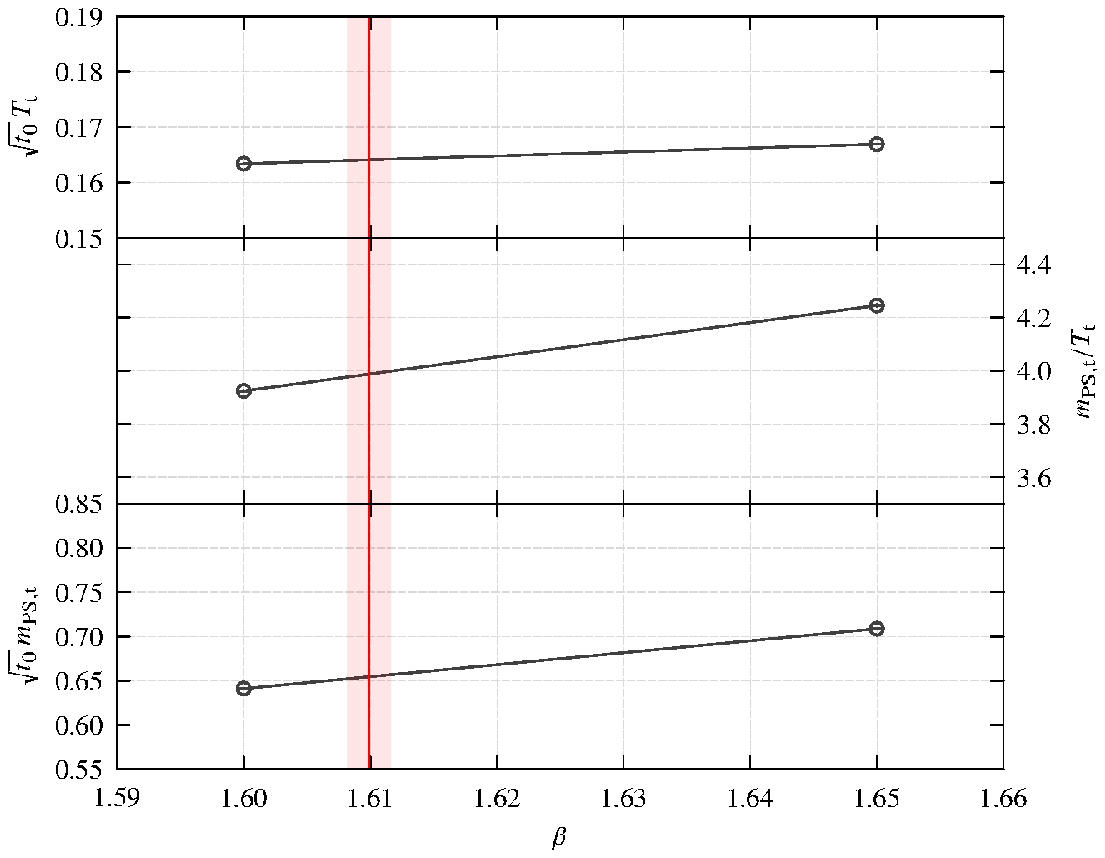}}
&
\scalebox{0.7}{\includegraphics{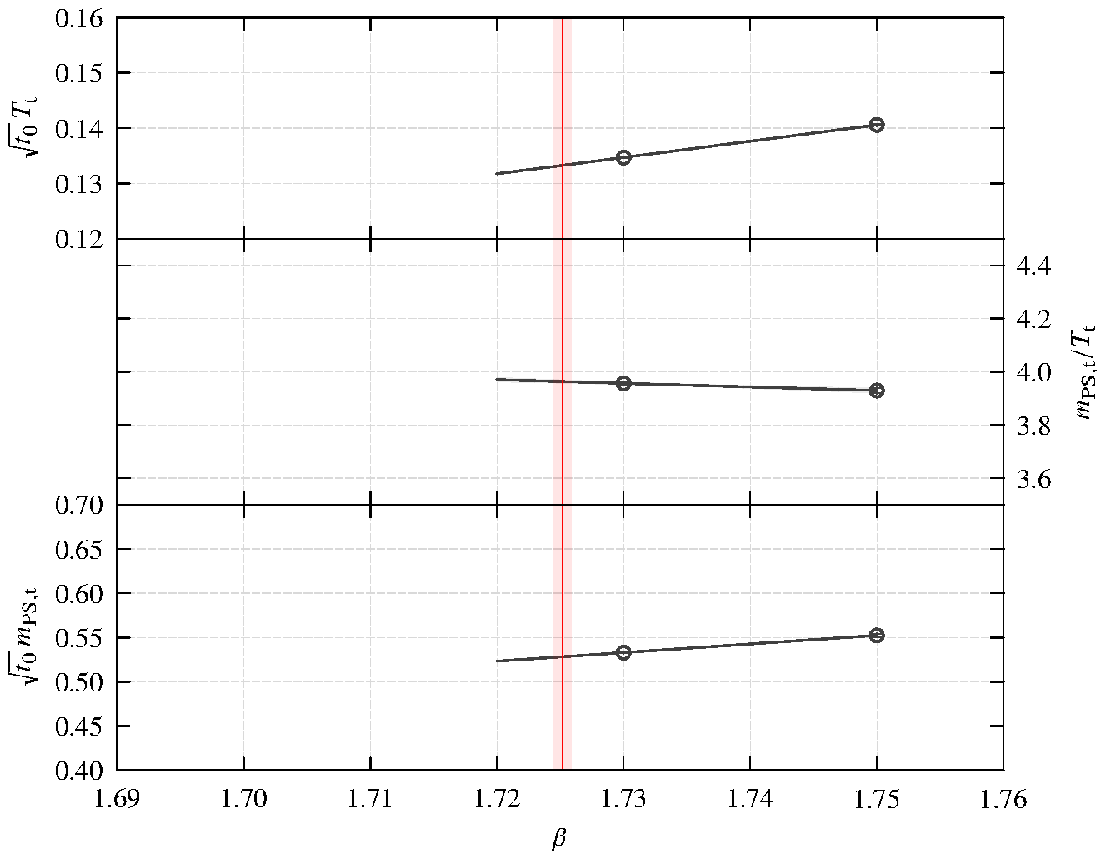}}
\\
\scalebox{0.7}{\includegraphics{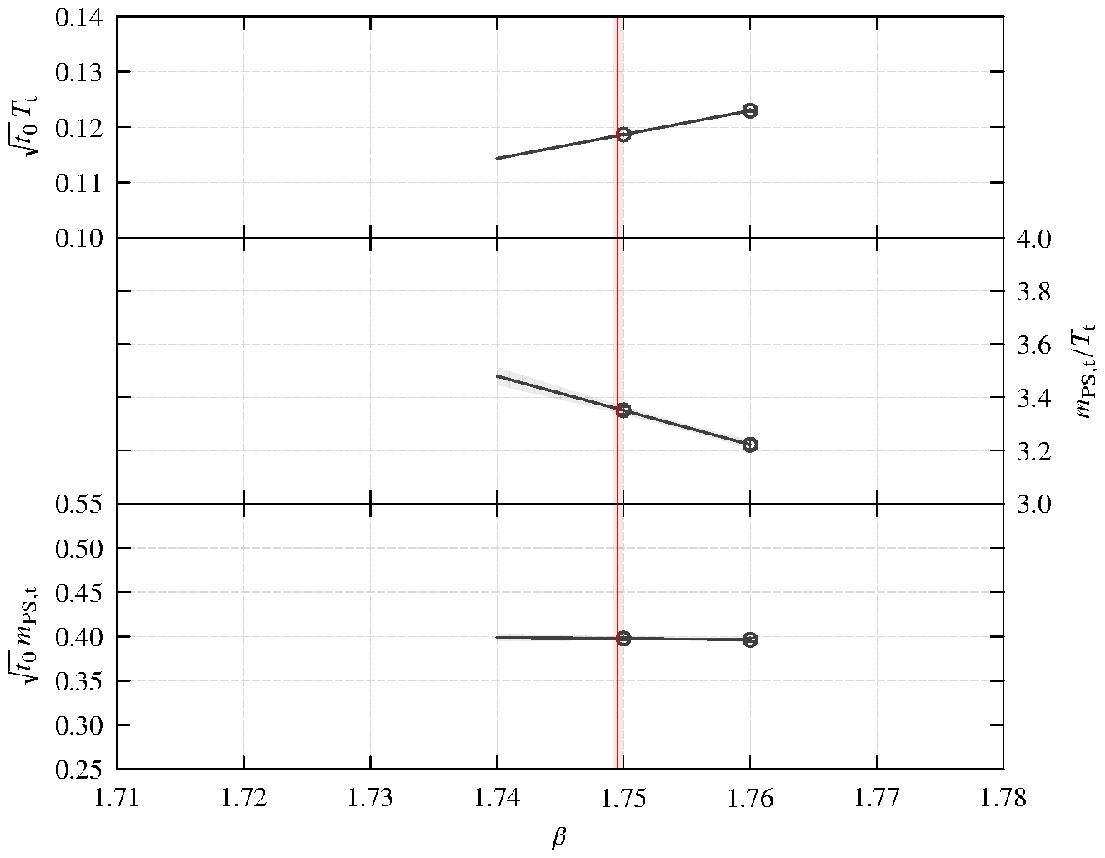}}
&
\scalebox{0.7}{\includegraphics{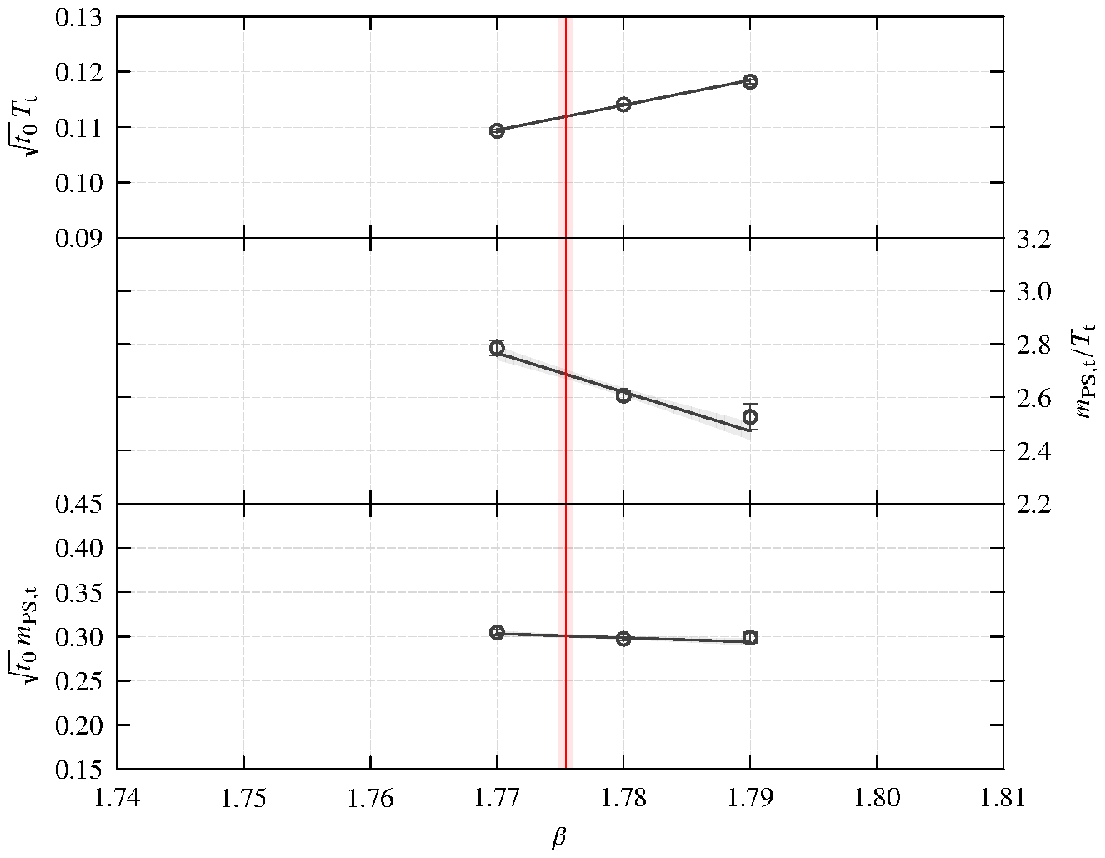}}
\end{tabular}
\end{center}
\caption{
$\sqrt{t_0}T$, $m_{\rm PS}/T$ and $\sqrt{t_0}m_{\rm PS}$ along the transition line projected on $\beta$ for $N_{\rm t}=4$(upper left), $6$(upper right), $8$(lower left) and $10$(lower right).
The vertical red line shows the location of the critical value of $\beta$ determined by the kurtosis intersection analysis.
}
\label{fig:hadron_transition_line}
\end{figure}

Before taking the continuum limit, let us summarize the dimensionless combination of the pseudo scalar meson mass $m_{\rm PS}$, the Wilson flow scale
$\sqrt{t_0}$ \cite{Luscher:2010iy} and the temperature $T$ along the transition line in Fig.~\ref{fig:hadron_transition_line}.
They are calculated by zero temperature simulations and their results are summarized in appendix \ref{sec:scale}.
The zero temperature simulation covers the parameter range of the critical points.
From an interpolation or a short extrapolation, one can obtain the critical value of the dimensionless quantities for each temporal size $N_{\rm t}$.
The actual numbers are summarized in Table~\ref{tab:continuum}.

\begin{table}[t!]
\caption{
The hadronic dimensionless quantities at the critical point for $N_{\rm t}$=4, 6, 8 and 10,
and their continuum extrapolation with various fitting range and fitting form.
}
\label{tab:continuum}
\begin{ruledtabular}
\begin{tabular}{llll}
$N_{\rm t}$
&
$\sqrt{t_0}m_{\rm PS,E}$
&
$\sqrt{t_0}T_{\rm E}$
&
$m_{\rm PS,E}/T_{\rm E}$
\\
\hline
4 &      0.6545(24) &     0.16409(13) &       3.987(12)\\
6 &      0.5282(12) &     0.13328(23) &      3.9630(63)\\
8 &      0.3977(19) &     0.11845(20) &       3.357(16)\\
10 &      0.3006(19) &     0.11193(29) &       2.687(18)\\
\hline
$\infty$ (fit) &0.0938(39) & 0.09970(37) &0.941(39) \\
$\infty$ (solve and quadratic) &0.1281(61) & -- &1.285(61) \\
$\infty$ (solve and cubic) &0.039(14) & -- &0.39(14) \\
\end{tabular}
\end{ruledtabular}
\end{table}

\begin{figure}[t]
\begin{center}
\begin{tabular}{cc}
\scalebox{0.7}{\includegraphics{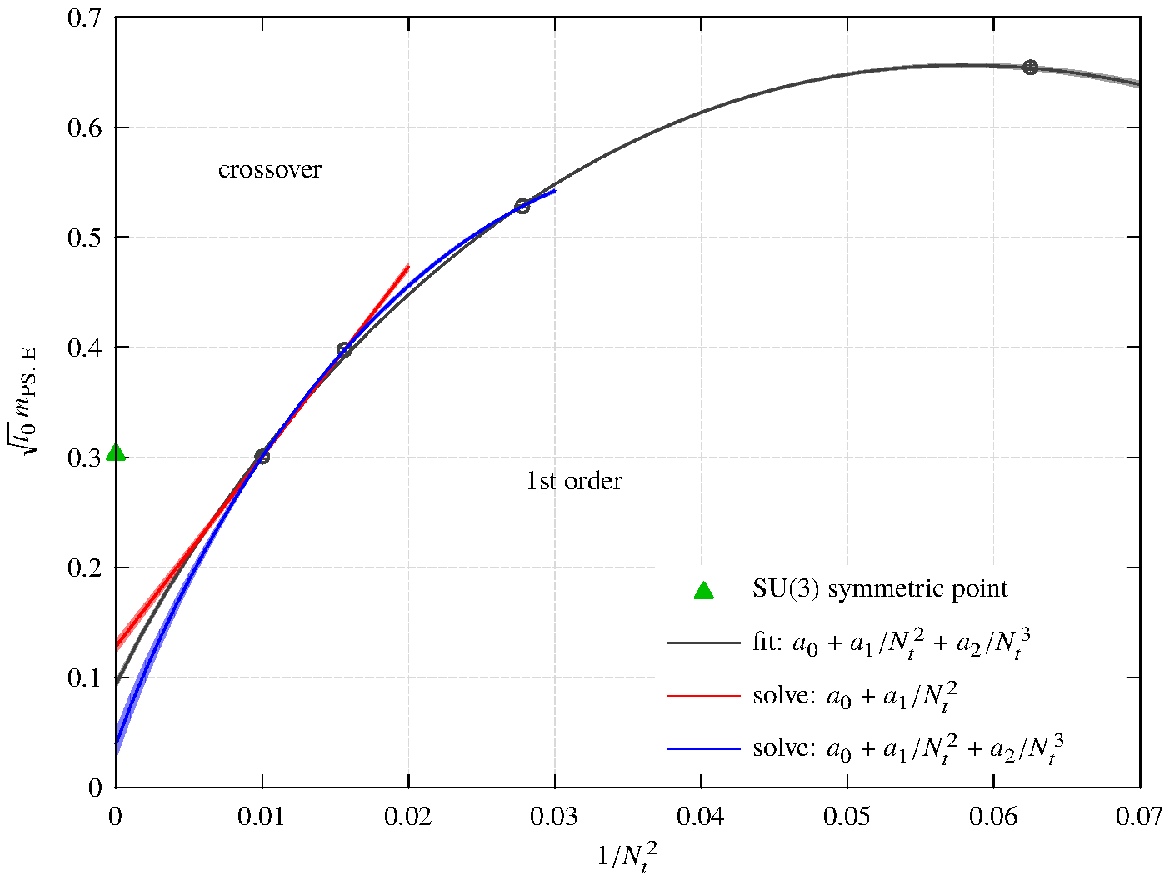}}
&
\scalebox{0.7}{\includegraphics{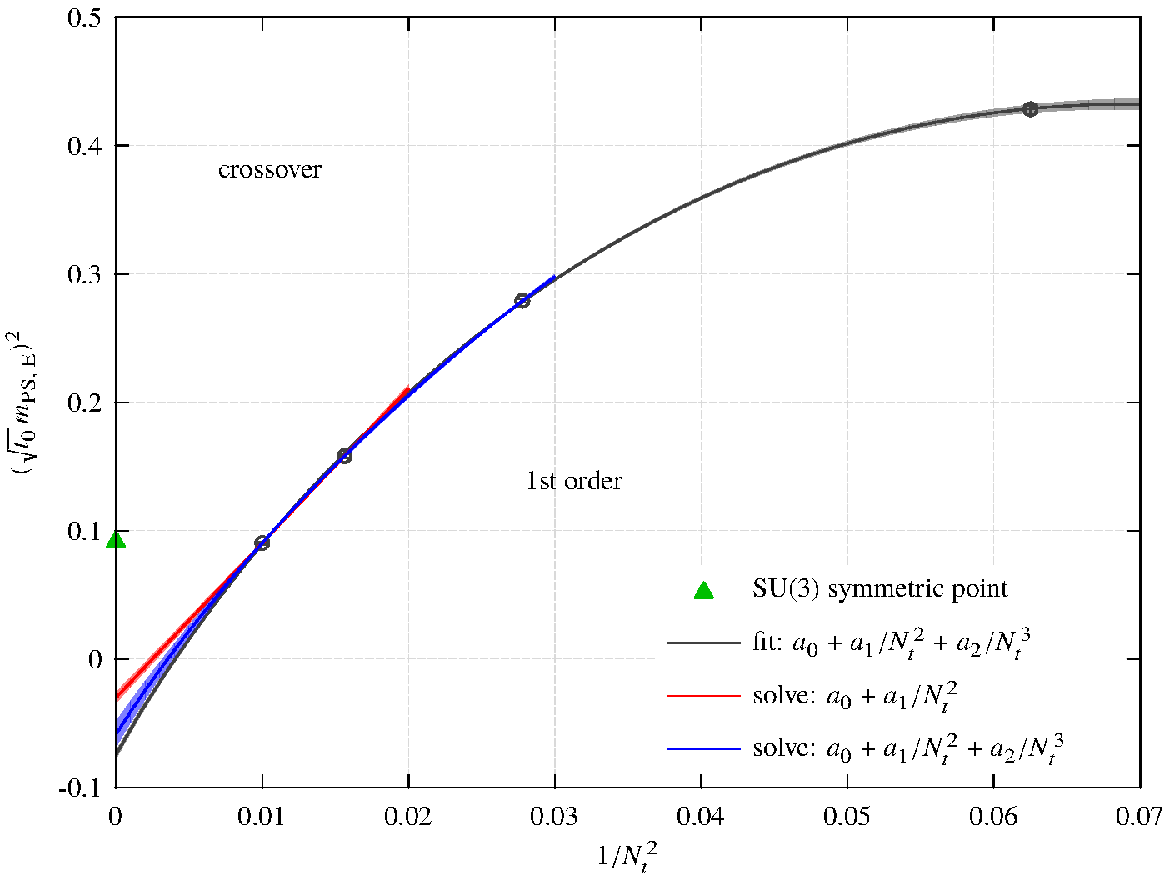}}
\\
\scalebox{0.7}{\includegraphics{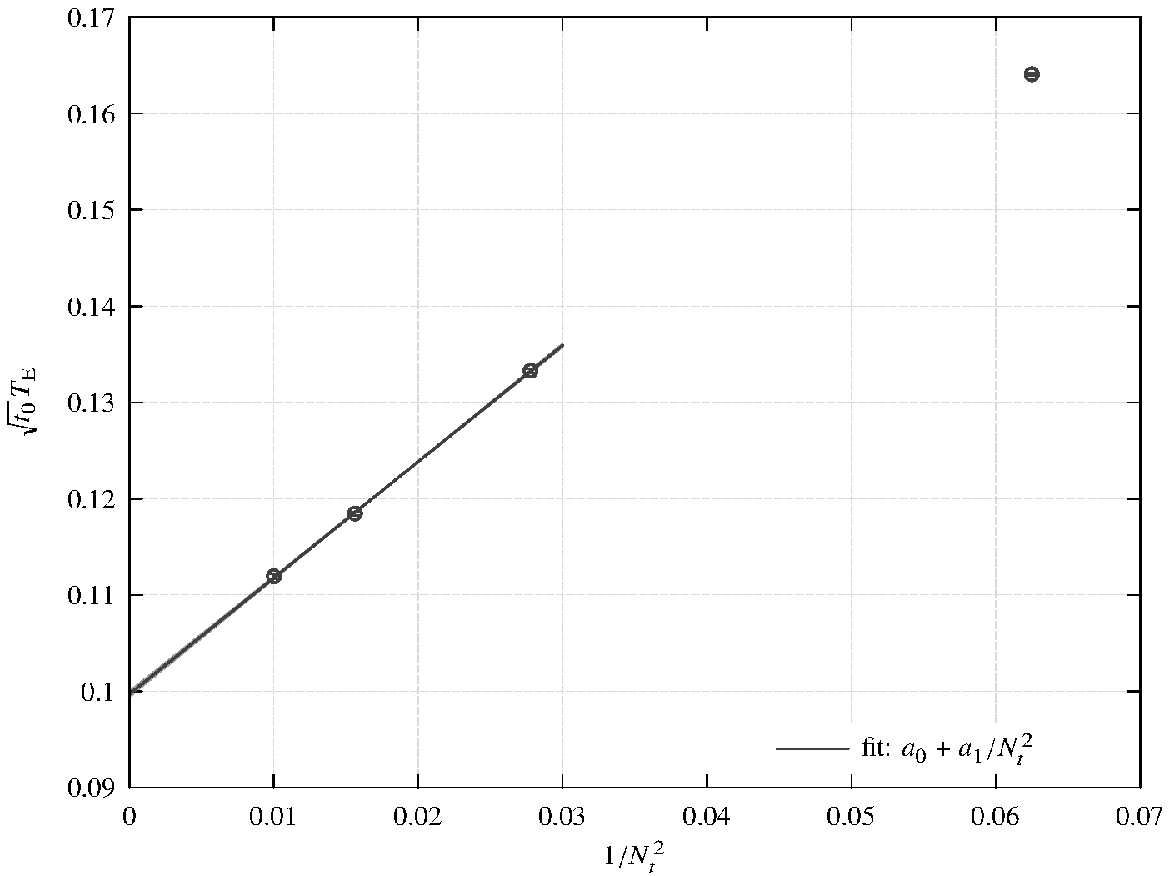}}
&
\scalebox{0.7}{\includegraphics{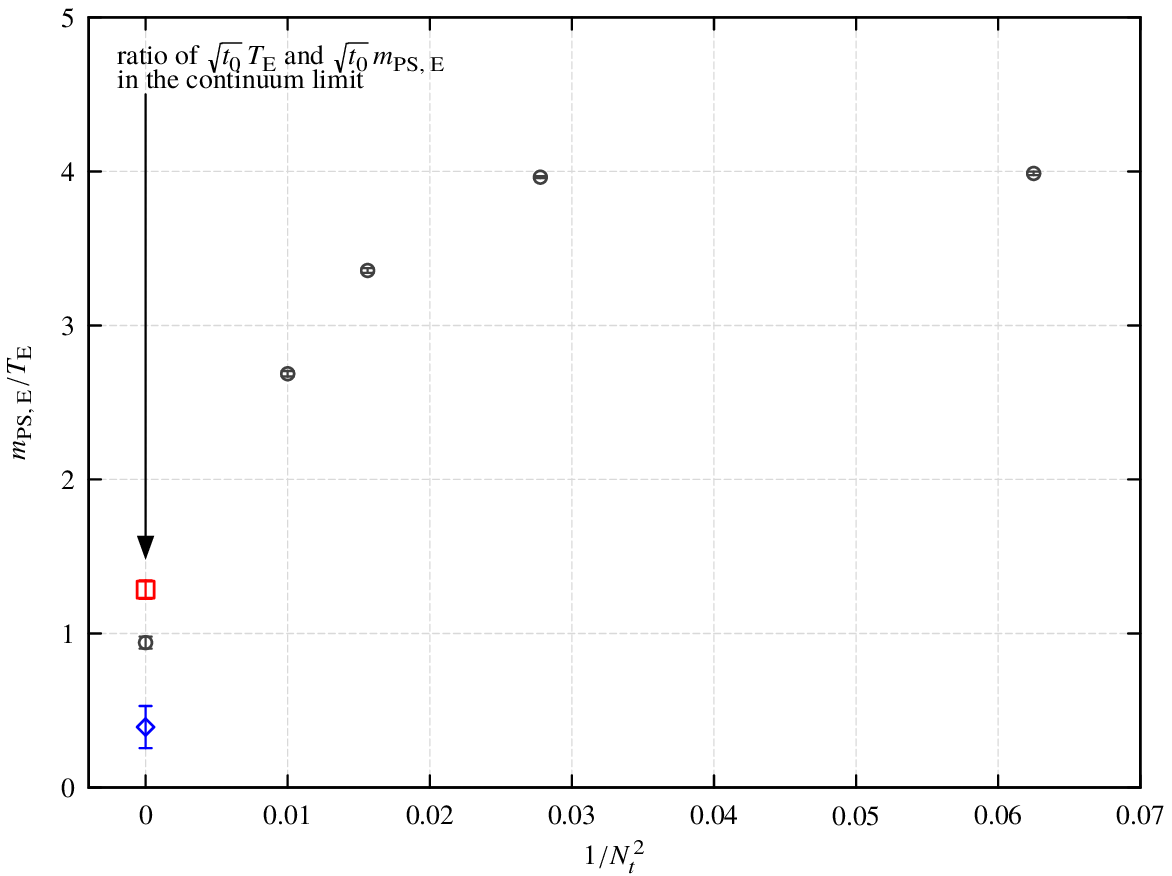}}
\end{tabular}
\end{center}
\caption{
Continuum extrapolation of the critical point
$\sqrt{t_0}m_{\rm PS,E}$ (upper-left),
$(\sqrt{t_0}m_{\rm PS,E})^2$ (upper-right),
$\sqrt{t_0}T_{\rm E}$ (lower-left)
and
$m_{\rm PS,E}/T_{\rm E}$ (lower-right).
Note that the continuum value of $m_{\rm PS,E}/T_{\rm E}$ in the lower-right panel is obtained by using the ratio of
$\sqrt{t_0}m_{\rm PS,E}$ and $\sqrt{t_0}T_{\rm E}$ in the continuum limit obtained in the upper-left and the lower-left panel respectively.
}
\label{fig:continuum}
\end{figure}

Finally, in Fig.~\ref{fig:continuum} (upper-left and lower-left panels) we show the continuum extrapolation of the critical pseudo scalar meson mass  
$\sqrt{t_0}m_{\rm PS,E}$ and the critical temperature $\sqrt{t_0}T_{\rm E}$ normalized by $\sqrt t_0$.
The latter shows a stable continuum extrapolation and we obtain $\sqrt{t_0}T_{\rm E}=0.09932(39)$.
The critical temperature in physical units is given by
$
T_{\rm E}
=
134(3)
{\rm MeV}
$
using the Wilson flow scale $1/\sqrt{t_0}=1.347(30)$GeV in Ref.~\cite{Borsanyi:2012zs}.
On the other hand $\sqrt{t_0}m_{\rm PS,E}$ shows significantly large scaling violation.
In the extrapolation procedure, we try some fitting forms including up to cubic correction term and examine the fitting range dependence.
As a result, their dependence turns out to be large as shown in Fig.~\ref{fig:continuum} (upper-left) and Table~\ref{tab:continuum}.
Furthermore we investigate the critical mass in terms of the quark mass like quantity, $(\sqrt{t_0}m_{\rm PS,E})^2\propto m_{\rm q}$ in Fig.~\ref{fig:continuum} (upper-right).
The result shows the large scaling violation as well and
extrapolates to a negative value in the continuum limit.
The inconsistency of the continuum value for $\sqrt{t_0}m_{\rm PS,E}$ and $(\sqrt{t_0}m_{\rm PS,E})^2$, in particular their signature, indicates that
a part of our data with $N_{\rm t}=4$--$10$ may not be in the scaling region.
Therefore, here we conservatively quote an upper bound of the critical value
$\sqrt{t_0}m_{\rm PS,E}\lesssim0.13$.
The value of the upper limit is taken from the maximum continuum value among all the fits we did.
In physical units, this bound is
$m_{\rm PS,E}
\lesssim
170
{\rm MeV}
$.
This upper bound is much smaller than our previous estimate ($\sim$300MeV) \cite{Jin:2014hea}, the reason being 
that the latest point at $N_{\rm t}=10$ (see Fig.~\ref{fig:continuum}) bends down toward the continuum extrapolation.

For future references, we address the continuum extrapolation of $m_{\rm PS,E}/T_{\rm E}$.
As shown in Fig.~\ref{fig:continuum} (lower-right), the lattice cut off dependence is quite large and
in fact the continuum extrapolation was not smoothly taken.
Therefore, instead of performing a direct continuum extrapolation of $m_{\rm PS,E}/T_{\rm E}$,
we take a ratio of the two values of
$\sqrt{t_0}m_{\rm PS,E}$
and 
$\sqrt{t_0}T_{\rm E}$
at the continuum limit and then obtain the upper bound
$m_{\rm PS,E}/T_{\rm E}\lesssim1.3$.

\section{Summary and outlook}
\label{sec:summary}

We carried out the large scale simulations for $N_{\rm t}=10$ and partly $N_{\rm t}=8$ by using the Wilson-type fermions. This is an extension of the previous work at $N_{\rm t}=4$, $6$ and $8$\cite{Jin:2014hea}.
We observed that the value of kurtosis at the crossing point tends to be larger as $N_{\rm t}$ increases.
To resolve the issue we derive and apply the modified formula to the kurtosis intersection analysis.
By using the formula the critical point is determined with assuming 3D Z$_2$ universality class.
We estimate the upper bound of the critical point and its temperature as
\begin{eqnarray}
m_{\rm PS,E}
&\lesssim&
170
{\rm MeV},
\\
T_{\rm E}
&=&
134(3)
{\rm MeV},
\\
m_{\rm PS,E}/T_{\rm E}
&\lesssim&
1.3.
\end{eqnarray}
Note that the continuum extrapolation significantly dominates the systematic error thus we compromise to quote the upper bound of $m_{\rm PS,E}$.
Since we are using the value of $c_{\rm sw}$ at very low $\beta$ which is out of the interpolation range \cite{Aoki:2005et},
it may be possible that the O($a$) improvement program is not  properly working in our parameter region.
In oder to control the lattice cutoff effect,
one can straightforwardly extend the temporal lattice size $N_{\rm t}$ but its cost is very demanding.
Another possibility may be to re-do the same calculation with a different lattice action,
say a different gauge action but the same/similar Wilson-type fermions.
And then one can perform a combined fit with an additional estimate of the critical point.

Our estimate of the upper bound of $m_{\rm PS,E}$ is larger than
that obtained by the staggered type fermions \cite{Bazavov:2017xul}, $m_{\rm PS,E}\lesssim50$ MeV.
Note that, however, our upper bound is derived from the existence of the critical point as an edge of the 1st oder phase transition while
the estimate of the smeared staggered study was based on its absence.
For $m_{\rm PS,E}/T_{\rm E}$, our bound is consistent with
the result of the standard staggered fermions \cite{deForcrand:2007rq,deForcrand:2017cgb}, $m_{\rm PS,E}/T_{\rm E}=0.37$.

Although our results of Wilson-type fermions is consistent with that of staggered-type fermions,
it is premature to conclude that the universality is confirmed.
In future as errors reduce a discrepancy may appear.
As seen above, the Wilson-type fermion is suffering from the large cut off effects, on the other hand,
the staggered fermions with the odd flavors may have a trouble in
the chiral regime at finite lattice spacing, namely the rooting issue.
Thus before studying $N_{\rm f}=3$ QCD intensively,
it is useful to study the universality for $N_{\rm f}=4$ QCD \cite{deForcrand:2017cgb} where there is no rooting issue and one can purely discuss the universality issue.
We are planning to study $N_{\rm f}=4$ QCD with Wilson-type fermions.

\section*{Acknowledgements}
This research used computational resources of 
HA-PACS and COMA provided by Interdisciplinary Computational Science Program in Center for Computational Sciences at University of Tsukuba,
System E at Kyoto University through the HPCI System Research project (Project ID:hp150141),
PRIMERGY CX400 tatara at Kyushu University
and
HOKUSAI GreatWave (project ID:G16016) at RIKEN.
This work is supported by JSPS KAKENHI Grant Numbers 26800130,
FOCUS Establishing Supercomputing Center of Excellence
and Kanazawa University SAKIGAKE Project.
This research used resources of the Argonne Leadership Computing Facility, which is a DOE Office of Science User Facility supported under Contract DE-AC02-06CH11357.

\appendix

\section{Wilson flow scale and pseudo scalar meson mass at zero temperature}
\label{sec:scale}

Simulation parameters, results for mass of pseudo-scalar meson $am_{\rm PS}$, and Wilson flow scale parameter $\sqrt{t_0}/a$
 are summarized in Tables~\ref{tab:scale160} and \ref{tab:scale176}. 
Result of following combined fit is given in Table~\ref{tab:scalefit},
\begin{eqnarray}
\label{eq:scalefit-1}
 (am_{\rm PS})^2 &=& a_1 \left({1\over \kappa} - {1 \over \kappa_{\rm c}}\right) + a_2 \left({1\over \kappa} - {1\over \kappa_{\rm c}}\right)^2 \,, \\
 \label{eq:scalefit-2}
{\sqrt{t_0} \over a} &=& b_0 + b_1 \left({1\over \kappa} - {1 \over \kappa_{\rm c}}\right) + b_2 \left({1\over \kappa} - {1 \over \kappa_{\rm c}}\right)^2\,.
\end{eqnarray}

\begin{table}[h]
\caption{\label{tab:scale160}
Simulation parameters, $\kappa$, $N_{\rm s}$, $N_{\rm t}$ and $\sqrt{t_0}/a$ and $am_{\rm PS}$ at $\beta=1.60$--$1.75$.}
\begin{ruledtabular}
\begin{tabular}{cccc|ll}
$\beta$ & $\kappa$ & $N_{\rm s}$ & $N_{\rm t}$ & $\sqrt{t_0}/a$ &  $am_{\rm PS}$  \\
\colrule
1.60&
  0.143000 &   12 &   24 &    0.650783(71) &     1.02752(71) \\
 & 0.143446 &   12 &   24 &    0.653722(72) &     0.98078(68) \\
 & 0.144000 &   12 &   24 &    0.658485(90) &     0.91122(74) \\
 & 0.145000 &   12 &   24 &     0.67160(15) &      0.7516(13) \\

\colrule
1.65&
  0.140000 &   12 &   24 &    0.659353(65) &     1.17770(72) \\
 & 0.141240 &   12 &   24 &     0.66818(11) &     1.06023(80) \\
 & 0.142000 &   12 &   24 &     0.67631(11) &     0.96934(88) \\
 & 0.143000 &   12 &   24 &     0.69391(21) &      0.8111(14) \\

\colrule
1.70&
  0.137100 &   12 &   24 &    0.673734(84) &     1.28494(85) \\
 & 0.137600 &   12 &   24 &     0.67752(11) &     1.24623(92) \\
 & 0.138100 &   12 &   24 &     0.68124(11) &     1.19924(87) \\
 & 0.138250 &   12 &   24 &     0.68277(12) &     1.18640(76) \\
 & 0.138610 &   12 &   24 &     0.68608(12) &     1.15202(69) \\
 & 0.140000 &   16 &   32 &    0.705367(99) &     0.99132(64) \\
 & 0.141000 &   16 &   32 &     0.73207(14) &      0.8243(15) \\
 & 0.141200 &   16 &   32 &     0.74115(22) &     0.77591(83) \\
 & 0.141456 &   16 &   32 &     0.75598(23) &     0.70619(86) \\
 & 0.141800 &   16 &   32 &      1.1510(12) &      0.4887(24) \\

\colrule
1.73&
  0.139000 &   12 &   24 &     0.73453(26) &     0.96412(97) \\
 & 0.139500 &   12 &   24 &     0.75087(27) &      0.8833(11) \\
 & 0.140000 &   16 &   32 &     0.77484(31) &      0.7787(11) \\
 & 0.140334 &   16 &   32 &     0.79915(38) &     0.68974(85) \\
 & 0.140435 &   16 &   32 &     0.80879(42) &     0.65851(99) \\
 & 0.140500 &   16 &   32 &     0.81630(36) &     0.63650(93) \\
 & 0.141000 &   16 &   32 &      0.9391(21) &      0.3306(45) \\

\colrule
1.75&
  0.139000 &   12 &   24 &     0.79055(42) &     0.82237(93) \\
 & 0.139500 &   12 &   24 &     0.82671(76) &      0.7017(15) \\
 & 0.139529 &   16 &   32 &     0.82959(42) &     0.69470(91) \\
 & 0.139669 &   16 &   32 &     0.84360(53) &      0.6569(11) \\
 & 0.139700 &   16 &   32 &     0.84799(45) &     0.64517(96) \\
 & 0.139850 &   16 &   32 &     0.86861(51) &     0.59293(89) \\
 & 0.140000 &   16 &   32 &      0.8917(14) &      0.5355(29) \\
 & 0.140242 &   16 &   32 &      0.9508(10) &      0.4176(18) \\
\end{tabular}
\end{ruledtabular}
\end{table}

\begin{table}[h]
\caption{\label{tab:scale176}
Simulation parameters, $\kappa$, $N_{\rm s}$, $N_{\rm t}$ and $\sqrt{t_0}/a$ and $am_{\rm PS}$ at $\beta=1.76$--$1.79$.}
\begin{ruledtabular}
\begin{tabular}{cccc|ll}
$\beta$&$\kappa$ & $N_{\rm s}$ & $N_{\rm t}$ & $\sqrt{t_0}/a$ &  $am_{\rm PS}$  \\
\colrule
1.76&
  0.139000 &   16 &   32 &     0.83107(33) &     0.73691(77) \\
 & 0.139500 &   16 &   32 &     0.88650(51) &     0.59667(93) \\
 & 0.139800 &   16 &   32 &     0.94019(98) &      0.4839(14) \\
 & 0.139850 &   16 &   32 &      0.9530(12) &      0.4567(17) \\
 & 0.139950 &   16 &   32 &      0.9823(13) &      0.4060(14) \\

\colrule
1.77&
  0.137100 &   12 &   24 &     0.77014(39) &      1.0040(12) \\
 & 0.137670 &   12 &   24 &     0.79076(35) &     0.91999(86) \\
 & 0.138500 &   12 &   24 &     0.83773(53) &      0.7675(12) \\
 & 0.138700 &   12 &   24 &     0.85652(79) &      0.7172(18) \\
 & 0.138903 &   16 &   32 &     0.87524(52) &     0.66902(80) \\
 & 0.139000 &   16 &   32 &     0.88795(57) &     0.63966(81) \\
 & 0.139653 &   16 &   32 &      1.0096(13) &      0.4063(14) \\
 & 0.139750 &   16 &   32 &      1.0447(17) &      0.3528(20) \\
 & 0.139850 &   16 &   32 &      1.0903(34) &      0.2851(36) \\
 & 0.139900 &   16 &   32 &      1.1163(52) &      0.2433(49) \\

\colrule
1.78&
  0.139356 &   16 &   32 &      1.0299(67) &      0.4079(21) \\
 & 0.139500 &   16 &   32 &      1.0910(20) &      0.3340(16) \\
 & 0.139600 &   16 &   32 &      1.1306(18) &      0.2729(16) \\
 & 0.139650 &   16 &   32 &      1.1615(28) &      0.2332(22) \\

\colrule
1.79&
  0.139000 &   16 &   32 &      1.0586(22) &      0.4253(34) \\
 & 0.139200 &   16 &   32 &      1.1158(18) &      0.3419(14) \\
 & 0.139300 &   16 &   32 &      1.1516(21) &      0.2903(21) \\
 & 0.139400 &   16 &   32 &      1.2019(28) &      0.2270(28) \\
\end{tabular}
\end{ruledtabular}
\end{table}

\begin{table}[h]
\caption{\label{tab:scalefit} Fit results to  (\ref{eq:scalefit-1}) and (\ref{eq:scalefit-2})  for critical hopping parameter $\kappa_{\rm c}$ and 
coefficients for  pseudo scalar meson mass $am_{\rm PS}$ and Wilson flow parameter $\sqrt{t_0}/a$ for $\beta=1.60$ to 1.79. }
\begin{ruledtabular}
\begin{tabular}{c|ccccccc|c}
$\beta$ & $\kappa_{\rm c}$ & $a_1$ & $a_2$ &$b_0$ &  $b_1$ & $b_2$ & $\chi^2/\dof$  & fit range \\
            &                 &          &         &        &            &         &                  &  $\kappa>$  \\
\colrule
 1.60 &         0.146763(36) &        7.61(17) &       -9.57(74) &      0.7064(13) &      -0.516(14) &       1.149(46)  &   2.18  & 0.1430 \\
 1.65 &         0.145000(29) &        7.60(11) &       -8.01(35) &      0.7409(12) &     -0.5967(99) &       1.079(25)  &  27.55 & 0.1400  \\
 1.70 &         0.142488(22) &       11.02(25) &      -25.2(1.7) &      0.8360(26) &      -1.926(37) &        7.05(18) &       2.40 & 0.1400 \\
 1.73 &        0.1411461(64) &       15.62(34) &      -98.1(7.5) &      0.9944(42) &       -8.17(21) &       83.2(3.2) &       0.90 & 0.1403 \\
 1.75 &        0.1405835(88) &       10.55(25) &      -29.0(3.7) &      1.0504(41) &       -6.59(15) &       46.3(1.8) &       2.53 & 0.1395 \\
 1.76 &         0.140276(17) &       10.56(62) &         -40(12) &       1.109(12) &       -9.06(65) &          87(11) &       1.38 & 0.1395 \\
 1.77 &        0.1400313(72) &        9.00(27) &      -24.2(4.2) &      1.1814(69) &      -10.85(46) &      100.3(6.5) &       0.34 & 0.1390 \\
 1.78 &         0.139792(12) &        7.60(69) &          -8(24) &       1.233(14) &      -11.2(1.9) &         103(71) &       3.36 & 0.1390 \\
 1.79 &         0.139555(13) &        6.52(52) &          -7(17) &       1.287(11) &      -12.04(92) &         142(24) &       0.92 & 0.1390 \\
\end{tabular}
\end{ruledtabular}
\end{table}


\begin{thebibliography}{99}

\bibitem{Brown:1990ev} 
  F.~R.~Brown, F.~P.~Butler, H.~Chen, N.~H.~Christ, Z.~h.~Dong, W.~Schaffer, L.~I.~Unger and A.~Vaccarino,
  Phys.\ Rev.\ Lett.\  {\bf 65}, 2491 (1990).
  
  
\bibitem{Pisarski:1983ms} 
  R.~D.~Pisarski and F.~Wilczek,
  Phys.\ Rev.\ D {\bf 29}, 338 (1984).
  
\bibitem{Gavin:1993yk} 
  S.~Gavin, A.~Gocksch and R.~D.~Pisarski,
  Phys.\ Rev.\ D {\bf 49}, R3079 (1994)
  [hep-ph/9311350].
  

\bibitem{Schmidt:2017bjt} 
  C.~Schmidt and S.~Sharma,
  arXiv:1701.04707 [hep-lat].

\bibitem{Ding:2017giu} 
  H.~T.~Ding,
  PoS LATTICE {\bf 2016}, 022 (2017)
  [arXiv:1702.00151 [hep-lat]].
  
\bibitem{Aoki:1998gia} 
  S.~Aoki {\it et al.} [JLQCD Collaboration],
  Nucl.\ Phys.\ Proc.\ Suppl.\  {\bf 73}, 459 (1999)
  [hep-lat/9809102].
  
\bibitem{Karsch:2001nf} 
  F.~Karsch, E.~Laermann and C.~Schmidt,
  Phys.\ Lett.\ B {\bf 520}, 41 (2001)
  [hep-lat/0107020].
    
\bibitem{deForcrand:2006pv} 
  P.~de Forcrand and O.~Philipsen,
  JHEP {\bf 0701}, 077 (2007)
  [hep-lat/0607017].

\bibitem{Smith:2011pm}
  D.~Smith and C.~Schmidt,
  PoS LATTICE {\bf 2011} (2011) 216
  [arXiv:1109.6729 [hep-lat]].
  
\bibitem{deForcrand:2007rq} 
  P.~de Forcrand, S.~Kim and O.~Philipsen,
  PoS LAT {\bf 2007}, 178 (2007)
  [arXiv:0711.0262 [hep-lat]].

\bibitem{Endrodi:2007gc} 
  G.~Endrodi, Z.~Fodor, S.~D.~Katz and K.~K.~Szabo,
  PoS LAT {\bf 2007}, 182 (2007)
  [arXiv:0710.0998 [hep-lat]].
  
\bibitem{Ding:2011du} 
  H.-T.~Ding, A.~Bazavov, P.~Hegde, F.~Karsch, S.~Mukherjee and P.~Petreczky,
  PoS LATTICE {\bf 2011}, 191 (2011)
  [arXiv:1111.0185 [hep-lat]].

\bibitem{Bazavov:2017xul} 
  A.~Bazavov, H.-T.~Ding, P.~Hegde, F.~Karsch, E.~Laermann, S.~Mukherjee, P.~Petreczky and C.~Schmidt,
  Phys.\ Rev.\ D {\bf 95}, no. 7, 074505 (2017)
  [arXiv:1701.03548 [hep-lat]].
  
\bibitem{Iwasaki:1996zt} 
  Y.~Iwasaki, K.~Kanaya, S.~Kaya, S.~Sakai and T.~Yoshie,
  Phys.\ Rev.\ D {\bf 54}, 7010 (1996)
  [hep-lat/9605030].

\bibitem{Jin:2014hea} 
  X.~Y.~Jin, Y.~Kuramashi, Y.~Nakamura, S.~Takeda and A.~Ukawa,
  Phys.\ Rev.\ D {\bf 91}, no. 1, 014508 (2015)
  [arXiv:1411.7461 [hep-lat]].

\bibitem{nakamuralattice2015} X.~-Y.~Jin, Y.~Kuramashi, Y.~Nakamura, S.~Takeda and A.~Ukawa,  PoS(Lattice 2015), 160 (2015).

\bibitem{Takeda:2016vfj} 
  S.~Takeda, X.~Y.~Jin, Y.~Kuramashi, Y.~Nakamura and A.~Ukawa,
  PoS LATTICE {\bf 2016}, 384 (2017)
  [arXiv:1612.05371 [hep-lat]].
  
\bibitem{Iwasaki:2011np} 
  Y.~Iwasaki,
  Report No. UTHEP-118 (1983),
  arXiv:1111.7054 [hep-lat].
 
\bibitem{Aoki:2005et} 
  S.~Aoki {\it et al.} [CP-PACS and JLQCD Collaborations],
  Phys.\ Rev.\ D {\bf 73}, 034501 (2006)
  [hep-lat/0508031].
  
\bibitem{Nakamura:2010qh} 
  Y.~Nakamura and H.~Stuben,
  PoS LATTICE {\bf 2010}, 040 (2010)
  [arXiv:1011.0199 [hep-lat]].
  
\bibitem{Clark:2006fx} 
  M.~A.~Clark and A.~D.~Kennedy,
  Phys.\ Rev.\ Lett.\  {\bf 98}, 051601 (2007)
  [hep-lat/0608015].
  
\bibitem{Ferrenberg:1988yz} 
  A.~M.~Ferrenberg and R.~H.~Swendsen,
  Phys.\ Rev.\ Lett.\  {\bf 61}, 2635 (1988).

\bibitem{Kuramashi:2016kpb} 
  Y.~Kuramashi, Y.~Nakamura, S.~Takeda and A.~Ukawa,
  Phys.\ Rev.\ D {\bf 94}, no. 11, 114507 (2016)
  [arXiv:1605.04659 [hep-lat]].

\bibitem{Luscher:2010iy} 
  M.~L\"uscher,
  JHEP {\bf 1008}, 071 (2010)
  Erratum: [JHEP {\bf 1403}, 092 (2014)]
  [arXiv:1006.4518 [hep-lat]].

\bibitem{Borsanyi:2012zs} 
  S.~Borsanyi {\it et al.},
  JHEP {\bf 1209}, 010 (2012)
  [arXiv:1203.4469 [hep-lat]].
    
\bibitem{deForcrand:2017cgb} 
  P.~de Forcrand and M.~D'Elia,
  PoS LATTICE {\bf 2016}, 081 (2017)
  [arXiv:1702.00330 [hep-lat]].
  
\end{thebibliography}
\end{document}